\documentclass[aps,pra,showpacs]{revtex4}
\usepackage{graphicx}
\usepackage{amsmath}

\def\<{\langle}
\def\>{\rangle}

\begin{document}

\title{One-party Quantum Error Correcting Codes for Unbalanced Errors: Principles and Application to
Quantum Dense Coding and Quantum Secure Direct Communications}
\author{Kai Wen$^{1}$ and Gui Lu Long$^{1,2,3}$ }
\affiliation{ $^1$ Key Laboratory For Quantum Information and Measurements and
Department of Physics, Tsinghua University,
Beijing 100084, China\\
$^2$ Key Laboratory for Atomic and Molecular NanoSciences, Tsinghua University,
Beijing
100084, China\\
$^3$ Center of Quantum Information Science and Technology, Tsinghua National
Laboratory For Information Science and Technology, Beijing 100084, China}
\date{28 January 2007 }

\begin{abstract}
In this article, we present the unbalanced quantum error correcting
codes(one-party-QECC), a novel idea for correcting unbalanced quantum errors. In some
quantum communication tasks using entangled pairs, the error distributions between two
parts of the pairs are unbalanced, and. one party holds the whole entangled pairs at
the final stage, and he or she is able to perform joint measurements on the pairs. In
this situation the proposed one-party-QECC can improve error correction by allowing a
higher tolerated error rate. We have established the general correspondence between
linear classical codes and the one-party-QECC, and we have given the general definition
for this type quantum error correcting codes.It has been shown that the one-party-QECC
can correct errors as long as the error threshold is not larger than 0.5. The
one-party-QECC works even for fidelity less than 0.5 as long as it is larger than 0.25.
We give several concrete examples of the one-party-QECC. We provide the applications of
one-party-QECC in quantum dense coding so that it can function in noisy channels. As a
result, a large number of quantum secure direct communication protocols based on dense
coding is also able to be protected by this new type of one-party-QECC.
\end{abstract}
\pacs{03.67.Hk,03.65.Ud,03.67.Dd,03.67.-a}

\maketitle

\section{Introduction}
\label{s_intro}

Quantum communication emerges as an important technology in communication. Its
key advantage lies in the unconditional security benefited from the principles
of quantum mechanics. As is well-known however, quantum system is more likely
to be affected by environment than classical system, and leads to phenomena
such as decoherence. Thus, how to efficiently build reliable quantum
communication through noisy quantum channels is one of the primary tasks for
scientists. Quantum error correcting codes(QECC) are such a technique aiming
towards this goal\cite{ShorCode, Good, Steane,fengkq,chenhao}. QECC make use of
redundant qubits to encode quantum information. To correct errors using QECC,
we measure the qubit system to obtain error syndromes and then choose
appropriate recovering operations. Entanglement purification protocol(EPP) is
another quantum error correction method to produce high fidelity entangled
pairs from low fidelity entangled pairs by sacrificing a number of
them\cite{EPP1,EPP2}. Both QECC and EPP play an important role in quantum
communication through noisy quantum channels. Especially, under assistance of
QECC and EPP, we are able to perform unconditional secure quantum key
distribution\cite{bdsw, LoChau,Shor, sixstate, twoway, practical_twoway,
pabshor, pab2way}.

Investigating quantum communication, we discover that the distribution of
quantum errors can be classified into two kinds: balanced and unbalanced
errors. Most quantum error correction methods are dealing with balanced errors.
In quantum communication especially using entangled pairs, there are two kinds
of qubits: flying qubits, which are transmitted from one party to the other,
and home qubits, which remain in the same party. Most quantum error correction
methods do not distinguish the errors between these two kinds of qubits and
their capabilities of correcting errors on both of them are identical. In this
case, we define these errors as balanced errors between flying and home qubits,
and these quantum error correction methods as balanced quantum error correction
methods.

For example, in quantum dense coding\cite{densecoding}, Alice and Bob first
share a sequence of two-qubit Einstein-Podolsky-Rosen(EPR) pairs, each in one
of the four Bell-basis states, namely,
\begin{equation}
\begin{split}
|\Phi^+\>&=\frac{1}{\sqrt{2}}(|00\>+|11\>),\\
|\Phi^-\>&=\frac{1}{\sqrt{2}}(|00\>-|11\>),\\
|\Psi^+\>&=\frac{1}{\sqrt{2}}(|01\>+|10\>),\\
|\Psi^-\>&=\frac{1}{\sqrt{2}}(|01\>-|10\>).\\
\end{split}
\end{equation}
The two qubits in every EPR pair are distributed between Alice and Bob. Then
Alice sends her qubits to Bob through the noisy channel and these qubits are
the flying qubits. The qubits initially in Bob's hand are home qubits because
they remain in the same place all through the protocol. After Bob received all
flying qubits, he can perform EPP to correct the errors introduced in the
process. In particular, there are three kinds of errors on qubits, namely,
bit-flip errors, phase-flip errors and both bit-flip and phase-flip errors. The
three kinds of errors are represented by acting three different Pauli operators
\begin{equation}
X=\left(\begin{array}{cc}\ 0 & \ 1 \\ \ 1 & \ 0\end{array}\right),
Z=\left(\begin{array}{cc}\ 1 & \ 0 \\ \ 0 & -1\end{array}\right),
Y=\left(\begin{array}{cc}\ 0 & -i \\ \ i & \ 0\end{array}\right),
\end{equation}
on the qubits respectively. The central purpose of EPP is to produce high fidelity
pairs from such low fidelity pairs by sacrificing a number of them. Because EPP does
not distinguish the errors on flying qubits and home qubits, he is able to correct
errors on both of them with the same capabilities. Therefore, EPP in quantum dense
coding is a kind of balanced quantum error correction method.

In this paper, we propose a novel kind of quantum error correcting codes, called
one-party quantum error correcting codes, and one-party-QECC for short. The unbalanced
errors are defined as those whose distribution between the flying and home qubits are
unbalanced. Practically, home qubits can be stored in some quantum memory with low
errors. They can be easily preserved in very high fidelity using conventional quantum
error correcting codes with low errors. On the contrary, the flying qubits are probably
affected by great channel noise. The high error rates on these qubits may sometimes
exceed the capability of conventional quantum error correction methods. The significant
difference of error distribution on the two kinds of qubits motivate us to propose a
kind of QECC, named one-party-QECC. The one-party-QECC that makes use of this feature
is proved in this paper to achieve much higher error correction capability.

The essential of the high capability of one-party-QECC lies in the quantum correlation
in the entanglements between flying and home qubits. It employs joint operations in the
final error detection and recovering process such as Bell measurements between them.
This feature leads to the requirement that in the final stage of the protocol, one of
the communication parties should hold both flying and home qubits. Thus he or she can
then easily apply Bell measurements to measure error syndromes without classically
informing the other party. Timely, many kinds of quantum communication satisfy this
requirement for our one-party-QECC, especially protocols based on quantum dense coding.

Quantum dense coding is one of the most important protocols that one-party-QECC is able
to apply. It is because in dense coding Alice's flying qubits and Bob's home qubits are
initially entangled and in the final stage, Bob obtains both kind of qubits. It is
reasonable that Bob's home qubits is of much higher fidelity than flying qubits all
through the protocol and thus the protocol contains unbalanced errors. We present in
this paper the application of one-party-QECC in quantum dense coding and show its
advantages in protecting message against very high initial and channel error rates.

Dense coding and its generalization \cite{densecoding,longliu} have extensive
application in quantum information processing. Based on quantum dense coding,
one-party-QECC can be applied to many more other protocols, for instance quantum secure
direct communication(QSDC) \cite{long0,bf,core,long1,cai1,wangc}. A QSDC communicates
secret messages directly through the quantum channel without first establishing a
secret key to encrypt them. Thus QSDC has the advantages of directly read-out of the
secret message by the legitimate user, and providing the eavesdropper only blind
results under any circumstance\cite{long1}. A large number of QSDC protocols use
entangled states and also make use of dense coding. So we also employ one-party-QECC in
this kind of QSDC protocols to enhance their performance. In some secret sharing
schemes, dense coding technique is also exploited, for example, in Ref.\cite{dengpla}.
In such case, the proposed one-party-QECC may also work.

In this paper, we first give two example one-party-QECC: the $[[6,2,1]]$ one-party-QECC
in section \ref{s_621} and the concatenated [[6,2,1]] one-party-QECC in section
\ref{s_con}. Then we give a general theory of one-party-QECC in section \ref{s_gen},
and elaborate on its properties. In addition, we apply one-party-QECC to the quantum
dense coding protocol and QSDC and demonstrate its advantages in quantum communication
in section \ref{s_densecoding} and \ref{s_qsdc}.

\section{A Simple $[[6,2,1]]$  QECC with joint measurements}
\label{s_621}

In this section, we give a concrete and simple example of the one-party-QECC. The
example is [[6,2,1]] one-party-QECC, which encodes 2 bit information into 3 EPR pairs
and protects the state against at most one bit-flip and one phase-flip errors. Here we
use double brackets to emphasize that the code is quantum. We describe the definition
and operations of the codes. Then based on the example, we discuss the restrictions and
application conditions of this code in quantum communication.

The motivation of designing the new one-party-QECC comes from quantum communication
using entangled pairs, namely, EPR pairs. Several protocols, including quantum dense
coding protocol,  quantum key distribution and quantum secure direct
communications\cite{long0,bf,core,cai1,long1,wangc} belong to such type of quantum
communication. Especially, due to the relationship between QECC and EPP\cite{bdsw},
previous work has proved that the fact that two-way classical communications in which
Alice and Bob compare the error syndromes and decide the next operations provides
higher capability of entanglement purification than one-way classical
communications\cite{twoway}. If we can further throw the classical communications in
the post-processing step,  we may obtain even higher capability. Indeed, this has been
achieved in this work.

Now we give the framework of the quantum communication that is suitable for our new
one-party-QECC. In a typical scheme of quantum communication using entangled pairs, the
errors distributed on both halves of the pairs are unbalanced: one half of the pairs
are usually kept in one party, called home qubits, and assumed to have no error, while
the other half are traveling, called flying qubits, through the noisy channels and may
err. Suppose the kind of quantum communication in which Bob obtains both qubit of an
EPR pair in the final stage, contrast to distributing the two qubits in a pair between
Alice and Bob in EPP. This kind of communication includes the protocols such as quantum
dense coding. In this kind of communication, when Bob obtains both qubit, he can make
use of joint operations, especially Bell measurements. Bell measurements can
successfully distinguish the four different Bell states. Bell measurements on a single
EPR pair can be described as measuring its $X_1 X_2$ and $Z_1 Z_2$ and each measurement
has two different outcomes. The combined 4 measurement results tell us about the state
of the EPR pairs as shown in Table \ref{bellmeasurement}. Using Bell measurements, one
can compare the difference between the travelling qubit which may be in error and the
local qubit which has no error, from the quantum correlation in the pair instead of
classical correlation using local operations and classical communications. This will
provide higher capability in error correction. Moreover, as the process of the
communication is different from EPP, we define this new kind of error correction
process using EPR pairs and Bell measurements as quantum error correcting codes with
joint measurements, or one-party-QECC for short.

\begin{table*}[page]
\caption{\label{bellmeasurement}Results of Bell measurements}
\begin{tabular}{ccc}
  \hline
  $X_1 X_2$ & $Z_1 Z_2$ & Bell state \\
  \hline
  \hline
  \ \ 1 & \ \ 1 & $|\Phi^+\>$ \\
  \hline
  \ \ 1 & $-1$ & $|\Psi^+\>$ \\
  \hline
  $-1$ & \ \ 1 & $|\Phi^-\>$ \\
  \hline
  $-1$ & $-1$ & $|\Psi^-\>$ \\
  \hline
\end{tabular}
\end{table*}

Following the above idea, we present a simple example of the one-party-QECC. We use EPR
pairs, instead of qubits, as the logical qubits. In the simplest classical error
correcting code -- the [3,1] error correcting code which encodes 1 bit information into
3 physical bits and protects 1 single bit-flip error in any of the 3 physical bits, the
logical $0$ and $1$ is encoded into $000$ and $111$ respectively. Therefore, we define
our logical states as the following:
\begin{equation}\label{62joint}
\begin{split}
|\overline{00}\>=|\Phi^+\>_{12}|\Phi^+\>_{34}|\Phi^+\>_{56},\\
|\overline{01}\>=|\Phi^-\>_{12}|\Phi^-\>_{34}|\Phi^-\>_{56},\\
|\overline{10}\>=|\Psi^+\>_{12}|\Psi^+\>_{34}|\Psi^+\>_{56},\\
|\overline{11}\>=|\Psi^-\>_{12}|\Psi^-\>_{34}|\Psi^-\>_{56}.\\
\end{split}
\end{equation}
Generally, any logical state can be represented as
\begin{equation}
\begin{split}
&a|\overline{00}\>+b|\overline{01}\>+c|\overline{10}\>+d|\overline{11}\>\\
 =&a|\Phi^+\>_{12}|\Phi^+\>_{34}|\Phi^+\>_{56}+ \\
  &b|\Phi^-\>_{12}|\Phi^-\>_{34}|\Phi^-\>_{56}+ \\
  &c|\Psi^+\>_{12}|\Psi^+\>_{34}|\Psi^+\>_{56}+ \\
  &d|\Psi^-\>_{12}|\Psi^-\>_{34}|\Psi^-\>_{56}. \label{logical3}
\end{split}
\end{equation}
In the definition of the logical states, we encode the four logical Bell states,
defined as the four logical states, in the left side of Eq.(\ref{62joint}) with 3
physical Bell states given in the right side of Eq.(\ref{62joint}) which contain 6
physical qubits with the subscripts from 1 to 6. Thus, we equivalently encode 2 logical
qubit into 6 physical qubits and name such type of one-party-QECC as the [[6,2]]
one-party-QECC.

\begin{figure*}[here]
\includegraphics[width=8cm]{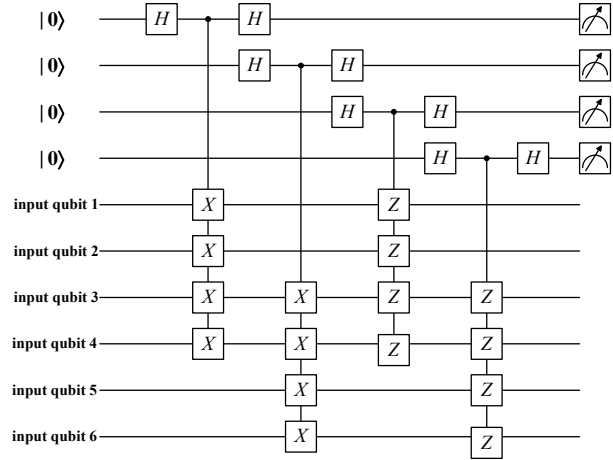}
\caption{Error detection quantum circuit of [[6,2,1]] one-party-QECC. The first four
ancillary qubits of initial $|0\>$ states are used to measure error syndromes $X_1 X_2
X_3 X_4$, $X_3 X_4 X_5 X_6$, $Z_1 Z_2 Z_3 Z_4$ and $Z_3 Z_4 Z_5 Z_6$ respectively. The
next six input qubits represent the input 3 EPR pairs.}\label{fig_321EPP}
\end{figure*}

As in any scheme of quantum error corrections, we should measure some operators
to get error syndromes in order to detect the error patterns. For the
definition of Eq.(\ref{62joint}) is induced by the classical $[3,1]$ error
correcting code, we can find the error syndromes in a similar way. In the
classical $[3,1]$ error correcting code, the error syndromes are measured by
$Z_1 Z_2$ and $Z_2 Z_3$. Thus, we measure
\begin{equation}
\begin{split}
g_1 &= Z_1 Z_2 Z_3 Z_4,\\
g_2 &= X_1 X_2 X_3 X_4,\\
g_3 &= Z_3 Z_4 Z_5 Z_6,\\
g_4 &= X_3 X_4 X_5 X_6,
\end{split}
\end{equation}
where the measurements act on the encoded state. We call the results of these 4
measurements as error syndromes which represent the error pattern of the EPR pairs.
According to the stabilizer coding theory\cite{stabilizer}, we have constructed an
error detection circuit of the [[6,2]] one-party-QECC as Fig.(\ref{fig_321EPP}). In
Fig.(\ref{fig_321EPP}), we use four ancillary qubits which are initially at $|0\>$
state to measure the error syndromes $Z_1 Z_2 Z_3 Z_4$, $X_1 X_2 X_3 X_4$, $Z_3 Z_4 Z_5
Z_6$ and $X_3 X_4 X_5 X_6$ of the input 6 qubits. Based on the measurements results of
the four test qubits, we can select proper operations to transform the corrupted states
to the correct ones.

\begin{table*}[page]
\caption{\label{errorsyndromes}Error syndromes of [[6,2,1]] one-party-QECC}
\begin{tabular}{cccccc}
  \hline
  $X_1 X_2 X_3 X_4$ & $X_3 X_4 X_5 X_6$ & $Z_1 Z_2 Z_3 Z_4$ & $Z_3 Z_4 Z_5 Z_6$ & Corresponding error & Correcting operations\\
  \hline
  \hline
  \ \ 1 & \ \ 1 & \ \ 1 & \ \ 1 & no error &  $I(Indentity)$ \\
  \hline
  \ \ 1 & \ \ 1 & \ \ 1 & $-1$ & bit 5 flip & $X_5$ \\
  \hline
  \ \ 1 & \ \ 1 & $-1$ & \ \ 1 & bit 1 flip & $X_1$ \\
  \hline
  \ \ 1 & \ \ 1 & $-1$ & $-1$ & bit 3 flip & $X_3$ \\
  \hline
  \ \ 1 & $-1$ & \ \ 1 & \ \ 1 & phase 5 flip & $Z_5$ \\
  \hline
  \ \ 1 & $-1$ & \ \ 1 & $-1$ & both bit and phase 5 flip & $Z_5 X_5$ \\
  \hline
  \ \ 1 & $-1$ & $-1$ & \ \ 1 & bit 1 and phase 5 flip & $Z_5 X_1$ \\
  \hline
  \ \ 1 & $-1$ & $-1$ & $-1$ & bit 3 and phase 5 flip & $Z_5 X_3$ \\
  \hline
  $-1$ & \ \ 1 & \ \ 1 & \ \ 1 & phase 1 flip & $Z_1$ \\
  \hline
  $-1$ & \ \ 1 & \ \ 1 & $-1$ & bit 5 and phase 1 flip & $Z_1 X_5$ \\
  \hline
  $-1$ & \ \ 1 & $-1$ & \ \ 1 & both bit and phase 1 flip & $Z_1 X_1$ \\
  \hline
  $-1$ & \ \ 1 & $-1$ & $-1$ & bit 3 and phase 1 flip & $Z_1 X_3$ \\
  \hline
  $-1$ & $-1$ & \ \ 1 & \ \ 1 & phase 3 flip & $Z_3$ \\
  \hline
  \ \ 1 & $-1$ & \ \ 1 & $-1$ & bit 5 and phase 3 flip & $Z_3 X_5$ \\
  \hline
  $-1$ & $-1$ & $-1$ & \ \ 1 & bit 1 and phase 3 flip & $Z_3 X_1$ \\
  \hline
  $-1$ & $-1$ & $-1$ & $-1$ & both bit and phase 3 flip & $Z_3 X_3$ \\
  \hline
\end{tabular}
\end{table*}

We now analyze the error model of the [[6,2]] one-party-QECC with some restriction and
present its capability in quantum error correction. As described in the framework of
quantum communication with unbalanced errors that is suitable for one-party-QECC, only
the flying qubits of the EPR pairs are transmitted through the noisy channel, while the
home qubits are kept in Bob's side, and stored in some quantum storages with
significantly low error rates compared to the quantum channels. Thus in the [[6,2]]
one-party-QECC, we assume that only one half qubit of the logical states are subject to
error. Particularly, in each logical state containing 3 EPR pairs, the 1st, 3rd and 5th
physical qubits are in Bob's hands, while the 2nd, 4th and 6th physical qubits are
transmitted through the noisy channel. Therefore, in our assumption, after the
transmission, only the 2nd, 4th and 6th physical qubits in each logical EPR pair may
have error, i.e., single bit-flip or phase-flip or both bit-flip and phase-flip error.
As a result, all error syndromes together with the error patterns and correcting
operations are shown in Table \ref{errorsyndromes}. From the table we can find that
these measurement are capable of detecting all errors in our example model of
one-party-QECC which is at most a single bit-flip and/or a single phase-flip on the
second half of EPR pairs. In conclusion, this kind of one-party-QECC can successfully
correct at most one bit-flip and one phase-flip errors occurring at the second half of
the 3 physical EPR pairs of each logical state and we name it as the [[6,2,1]]
one-party-QECC.

Furthermore, we define the logical operations of the [[6,2,1]] one-party-QECC which
enable Bob to encode the information into the logical states. The logical operations
are defined as
\begin{equation}
\begin{split}
\bar{X}_{\bar{1}}&=X_2 X_4 X_6 \\
\bar{Z}_{\bar{1}}&=Z_1 Z_2 Z_3 Z_4 Z_5 Z_6\\
\bar{X}_{\bar{2}}&=Z_1 Z_3 Z_5 \\
\bar{Z}_{\bar{2}}&=X_1 X_2 X_3 X_4 X_5 X_6
\end{split}
\end{equation}
The subscripts $\bar{1}$ and $\bar{2}$ in the left of the equations indicates which
logical qubits in Eq.(\ref{62joint}) the logical operators act on. These logical
operations are able to transform one encoded state into another encoded state, as shown
in Table \ref{62transform}. In all, we have successfully constructed an [[6,2,1]]
one-party-QECC, which encodes 2 bit information and are capable of correcting at most a
single bit-flip and a single phase-flip errors on the second halves of the physical EPR
pairs.

\begin{table*}[page]
\caption{\label{62transform}Transform between encoded states}
\begin{tabular}{ccccc}
  \hline
   & $\bar{X}_{\bar{1}}$ & $\bar{Z}_{\bar{1}}$ & $\bar{X}_{\bar{2}}$ & $\bar{Z}_{\bar{2}}$ \\
  \hline
  \hline
  $|\overline{00}\>$ & $\ \ |\overline{10}\>$ & $\ \ |\overline{00}\>$ & $\ |\overline{01}\>$ & $\ \ |\overline{00}\>$ \\
  \hline
  $|\overline{01}\>$ & $\ \ |\overline{11}\>$ & $\ \ |\overline{01}\>$ & $\ |\overline{00}\>$& $-|\overline{01}\>$  \\
  \hline
  $|\overline{10}\>$ & $\ \ |\overline{00}\>$ & $-|\overline{10}\>$ & $\ |\overline{11}\>$ & $\ \ |\overline{10}\>$ \\
  \hline
  $|\overline{11}\>$ & $\ \ |\overline{01}\>$ & $-|\overline{11}\>$ & $\ |\overline{10}\>$ & $-|\overline{11}\>$ \\
  \hline
\end{tabular}
\end{table*}

\section{Concatenating the [[6,2,1]] one-party-QECC}
\label{s_con}

In this section, we extend the [[6,2,1]] one-party-QECC into longer one-party-QECC by
concatenating it several rounds. The concatenated [[6,2,1]] one-party-QECC can thus be
used to encode longer sequences. Then we analyze the error threshold of the
concatenated [[6,2,1]] one-party-QECC.

The concatenating steps are described as the following: in the first round, Bob, the
final receiver, divides the final raw physical EPR sequence into groups, each with 3
consecutive EPR pairs. In each group, Bob applies the [[6,2,1]] one-party-QECC on the 3
EPR pairs. If the error correction of a group is successful, Bob obtains a correct
logical EPR pair, for the logical states of the one-party-QECC are actually the four
logical Bell states; otherwise he gets an error logical EPR pair. This logical EPR
sequence will be the building blocks for constructing logical EPR pairs for the next
round. Therefore, after applying the [[6,2,1]] one-party-QECC on all groups, Bob gets a
logical EPR sequence with $1/3$  original length consisting the logical states of the
[[6,2,1]] one-party-QECC on all groups. If the raw error rate is low enough, after the
first round, the error rate of the resulting logical EPR sequence will decrease.

\begin{figure}
\begin{center}
\includegraphics[width=14cm]{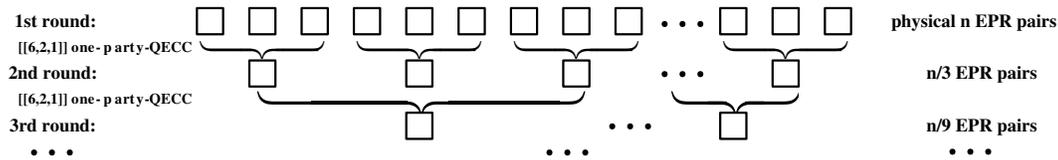}
\caption{The process of concatenating [[6,2,1]] one-party-QECC}\label{fig_cjEPP}
\end{center}
\end{figure}

In the second round, Bob divides the logical EPR sequence from the first round
into groups each with 3 consecutive logical EPR pairs. In particular, the
logical states of each group in the second round are defined as
\begin{equation}
\begin{split}
|\overline{00}\>^{(2)} &= |\overline{00}\>|\overline{00}\>|\overline{00}\>,\\
|\overline{01}\>^{(2)} &= |\overline{01}\>|\overline{01}\>|\overline{01}\>,\\
|\overline{10}\>^{(2)} &= |\overline{10}\>|\overline{10}\>|\overline{10}\>,\\
|\overline{11}\>^{(2)} &= |\overline{11}\>|\overline{11}\>|\overline{11}\>.\\
\end{split}
\end{equation}
Then he applies the [[6,2,1]] one-party-QECC on each group and obtains the correct
results as another logical EPR sequence in the same way as in the first round. Bob
repeats this procedure for several rounds. For the error rate is reduced in each round,
Bob will finally reach a very low error rate. When he obtains the required accuracy, he
finishes the error correcting procedure. A summary of the process of concatenating
[[6,2,1]] one-party-QECC is illustrated in Fig.(\ref{fig_cjEPP}).

Now we analyze the asymptotic error threshold of the concatenated [[6,2,1]]
one-party-QECC with nearly infinitely long initial raw physical EPR pairs. First we
assume that a quantum channel with the bit-flip and phase-flip error probabilities
which are both lower than $q_0$ in the beginning. After the $k$-th round of the
one-party-QECC, the bit-flip and phase-flip error rates becomes lower than $q_k$. Note
that, the error rate $q$ is defined as $q = \frac{t}{n}$, where $t$ is the number of
the bit-flip errors or the phase-flip errors and $n$ is the number of the physical
qubits that are transmitted through the quantum channels. In this type of quantum
communication and the error model, only the second half of each physical EPR pair is
transmitted through the quantum channel and subject to error; so the error rate $q$
also corresponds to the error rate of the physical EPR pairs. Thus the condition of the
success of the concatenated [[6,2,1]] one-party-QECC is that the error rate of $q_k$
should be less than $q_{k-1}$ for each round. Considering the $k$-th round, if there is
not more than one error out of the 3 EPR pairs in each group, the [[6,2,1]]
one-party-QECC can be successful in that group. Therefore, the total error rate of the
outcome logical EPR pair from 3 EPR pairs after the round is
\begin{equation}\label{EPP_result}
\begin{split}
q_k &= 1 - [(1-q_{k-1})^3 + 3 (1-q_{k-1})^2 q_{k-1}] \\
    &= (3-2q_{k-1})q^2_{k-1}
\end{split}
\end{equation}
The condition of the success of the concatenated [[6,2,1]] one-party-QECC requires that
$q_k < q_{k-1}$. Substituting Eq.(\ref{EPP_result}) into the inequality, we get
\begin{equation}\label{q_k}
q_{k-1}<0.5.
\end{equation}
The result means that if $q_{k-1}<0.5$, the $k$-th round will actually reduce the error
rate of the EPR sequence. Therefore, if the initial error rate, namely, $q_0$ is lower
than $50\%$, all rounds in the one-party-QECC will successfully reduce the error rates.

To show explicitly that the error rate in concatenating [[6,2,1] one-party-QECC indeed
converges to 0, we can find from Eq.(\ref{EPP_result}) that for any $0 < q_{k-1} < 1$,
\begin{equation}
\begin{split}
q_k &= (3-2q_{k-1})q^2_{k-1} \\
    &= 3q^2_{k-1} - 2q^3_{k-1} \\
    &< 3q^2_{k-1}.
\end{split}
\end{equation}
Thus, after $k$ rounds,
\begin{equation}
q_k < (\sqrt{3}q_0)^{2k}.
\end{equation}
Because we begin with $q_0 < 0.5$, we get $\sqrt{3}q_0 < 1$ and the error rate
decreases exponentially with $k$. As a result, if Bob selects a proper parameter $k$
according to $q_0$, he will obtain a total error rate of nearly 0, and the error
correction is successful. In conclusion, our new one-party-QECC can tolerate the bit
error rate up to 50\%. Compared to the upper bound of correctable bit error rate of
about 11\% in Calderbank-Shor-Steane codes\cite{Good, Steane}, one-party-QECC has
significant advantages in quantum error correction and are capable of correcting more
errors. The following sections will discuss the relation of the bounds between
different error correcting methods and one-party-QECC.

\section{General formalism of one-party-QECC}
\label{s_gen}

In this section, with the idea of the [[6,2,1]] one-party-QECC and the concatenated
[[6,2,1]] one-party-QECC in section \ref{s_621} and \ref{s_con}, we derive a general
formalism of our novel one-party-QECC. To achieve this, we establish a correspondence
between the classical linear codes and the one-party-QECC. We then present the logical
states, the measurements of the error syndromes and the logical operations of the
one-party-QECC from the correspondence to the classical linear codes. Finally, a bound
of the one-party-QECC is derived from the classical Gilbert-Vashamov
bound\cite{ClassicalGood} and a numerical result of the error-tolerating capacity of
the one-party-QECC is presented.

In section \ref{s_621}, we make use of the classical [3,1,1] code to construct the
[[6,2,1]] one-party-QECC. This leads us to investigate more general classical linear
codes and find its correspondence to the generalized one-party-QECC. Suppose there are
an $[n,k,t]$ classical linear code which encodes $k$ bits into $n$ bits and protects
the state against at most $t$ bit-flip errors, we can define an $[[2n,2k,t]]$
one-party-QECC which can encode $2k$ bits information into $n$ EPR pairs ($2n$ qubits)
and protect at most $t$ bit-flip errors and $t$ phase-flip errors on the second half of
the EPR pairs.

Note that we use the same error model as that in section \ref{s_621}. For in our
quantum communications, there are always two halves of the entangled pairs: one half
stay in Bob's hands throughout the whole process of the communications, which can be
stored in some quantum storage with very low errors and successfully protected by some
existing codes such as the CSS codes; the other half are transferred through the noisy
quantum channels but finally back to Bob's side, which will introduce much higher
errors. The existing error correcting codes may fail due to such high errors; however,
our new one-party-QECC can protect the half of the EPR pairs transmitted through the
quantum channels in this situation. We will show in the following text that they can
tolerate much higher error rates than existing codes. Consequently, we assume that in
our one-party-QECC, the errors can only happen in the second half of the physical EPR
pairs.

Now we present the detailed definition of the $[[2n,2k,t]]$ one-party-QECC. In details,
corresponding to an element
\begin{equation}\label{class_stab}
g^{cl}_i=Z^{cl}_{i_1} Z^{cl}_{i_2} \cdots Z^{cl}_{i_l}
\end{equation}
in the stabilizer of the classical code (here to avoid confusion, we use the
superscript of ``cl'' to indicate the classical operators), we define $2$ error
syndromes of the one-party-QECC measured by
\begin{equation}\label{joint_stab}
\begin{split}
g_{z,i}&=(Z_{2i_1 -1} Z_{2i_1})(Z_{2i_2 -1} Z_{2i_2})\cdots (Z_{2i_l-1} Z_{2i_l}), \\
g_{x,i}&=(X_{2i_1 -1} X_{2i_1})(X_{2i_2 -1} X_{2i_2})\cdots (X_{2i_l-1}
X_{2i_l})
\end{split}
\end{equation}
respectively. The subscript  $i_j$ for $\forall j = 1, 2, \cdots, l$ means the
operation is on the $i_j$-th EPR pair, and the index of qubits of the $i_j$-th EPR pair
are $2i_j -1$ and $2i_j$. Moreover, all $i, i_1, i_2, \cdots, i_l$ are of the same
values as those in Eq.(\ref{class_stab}) because of the correspondence between the
classical codes and the one-party-QECC. All error syndromes measured by the $Z$-type
operators $g_{z,i}$ completely give us the bit-flip error patterns, and all error
syndromes measured by the $X$-type operators $g_{x,i}$ completely give us the
phase-flip error patterns. Thus we can apply corresponding operations to correct the
bit-flip and phase-flip errors. Furthermore, similar to the [[6,2,1]] one-party-QECC,
we can create quantum circuit of error detection of $[[2n,2k]]$ one-party-QECC by
following stabilizer theory\cite{stabilizer}. For the classical stabilizers in
Eq.(\ref{class_stab}) are capable of discriminating any error which is caused by no
more than $t$ flipped bits, the stabilizers of the generalized one-party-QECC in
Eq.(\ref{joint_stab}) can successfully reveal any combination of at most $t$ bit-flip
and at most $t$ phase-flip errors occurring on the second halves of the physical EPR
pairs. Therefore, with the sufficient information of the error pattern, we can
completely correct the corrupted states; the one-party-QECC is called $[[2n,2k,t]]$
one-party-QECC.

After defining the error syndromes, the logical encoded states can be easily
worked out. However, there is another simple way: firstly, any logical state
can be represented as
\begin{equation}
|\overline{a_1 a_2;a_3 a_4;\cdots;a_{2k-1} a_{2k}}\>
\end{equation}
in terms of $a_i = 0, 1$, for $\forall i = 1, 2, \cdots, 2k$. The
representation of the logical state is divided into $k$ groups. In each group,
namely, the $i$-th group, there are 2 logical qubits $|\overline{a_{2i-1}
a_{2i}}\>$ in the total four different logical states, namely,
\begin{equation}
\begin{split}
|\overline{0_{2i-1} 0_{2i}}\> &= |\bar{\Phi}^+\>_{2i-1, 2i}, \\
|\overline{0_{2i-1} 1_{2i}}\> &= |\bar{\Phi}^-\>_{2i-1, 2i}, \\
|\overline{1_{2i-1} 0_{2i}}\> &= |\bar{\Psi}^+\>_{2i-1, 2i}, \\
|\overline{1_{2i-1} 1_{2i}}\> &= |\bar{\Psi}^-\>_{2i-1, 2i}. \\
\end{split}
\end{equation}
In order to transform between the logical encoded states, we remember that in
the classical codes, the logical state are transformed by the combination of
several logical bit-flip operators. The logical bit-flip operators of the
classical codes are defined as,
\begin{equation}\label{class_X}
\bar{X}^{cl}_i = X^{cl}_{i_1} X^{cl}_{i_2}\cdots X^{cl}_{i_l},
\end{equation}
for any $i = 1,2,\cdots,k$. Each operator can flip the corresponding logical
bit, namely,
\begin{equation}
\begin{split}
\bar{X}^{cl}_i 0_i &= 1_i, \\
\bar{X}^{cl}_i 1_i &= 0_i.
\end{split}
\end{equation}
As a result, with the correspondence between the classical linear codes and the
one-party-QECC, we define the logical operators for the one-party-QECC as,
\begin{equation}\label{joint_operator}
\begin{split}
\bar{X}_{2i-1} &= X_{2i_1} X_{2i_2} \cdots X_{2i_l}, \\
\bar{Z}_{2i-1} &= (Z_{2i_1-1} Z_{2i_1}) (Z_{2i_2-1} Z_{2i_2}) \cdots (Z_{2i_l-1} Z_{2i_l}), \\
\bar{X}_{2i} &= Z_{2i_1-1} Z_{2i_2-1} \cdots Z_{2i_l-1}, \\
\bar{Z}_{2i} &= (X_{2i_1-1} X_{2i}) (X_{2i_2-1} X_{2i_2}) \cdots (X_{2i_l-1}
X_{2i_l}).
\end{split}
\end{equation}
Again we require that the subscripts $i, i_1, i_2, \cdots, i_l$ are of the same
values to those in Eq.(\ref{class_X}). These logical operators act as the
bit-flip and phase-flip operators on the encoded states, i.e.,
\begin{equation}
\begin{split}
&\bar{X}_{2i-1} |\bar{0}\>_{2i-1} = |\bar{1}\>_{2i-1}, \\
&\bar{X}_{2i-1} |\bar{1}\>_{2i-1} = |\bar{0}\>_{2i-1}, \\
&\bar{Z}_{2i-1} |\bar{0}\>_{2i-1} = |\bar{0}\>_{2i-1}, \\
&\bar{Z}_{2i-1} |\bar{1}\>_{2i-1} = -|\bar{1}\>_{2i-1}, \\
&\bar{X}_{2i} |\bar{0}\>_{2i} = |\bar{1}\>_{2i}, \\
&\bar{X}_{2i} |\bar{1}\>_{2i} = |\bar{0}\>_{2i}, \\
&\bar{Z}_{2i} |\bar{0}\>_{2i} = |\bar{0}\>_{2i}, \\
&\bar{Z}_{2i} |\bar{1}\>_{2i} = -|\bar{1}\>_{2i}.
\end{split}
\end{equation}
Furthermore, in the classical codes, traditionally, it always holds that
\begin{equation}
\underbrace{\overline{00\cdots 0}}_{k} = \underbrace{00\cdots 0}_{n}
\end{equation}
and any logical states can be created by the combination of several logical bit-flip
operators acting on such logical zero state. Therefore, we begin with the logical zero
state of the one-party-QECC as
\begin{equation}\label{initial_state}
|\underbrace{\overline{00;00;\cdots;00}}_{k}\> = |\Phi^+\>_{12} |\Phi^+\>_{34}
\cdots |\Phi^+\>_{2n-1, 2n}.
\end{equation}
(note that the right side of the equation is the state of the physical qubits), and
employ the combination of the logical operators in Eq.(\ref{joint_operator}) to create
any desired encoded states of the $[[2n,2k]]$ one-party-QECC. In this way, we have
worked out both the logical states and the logical operations of the generalized
one-party-QECC.

In the end of this section, we estimate the asymptotic tolerable error rate of the
generalized one-party-QECC. Intuitively, from the correspondence between the classical
linear codes and the generalized one-party-QECC, we can say that one-party-QECC can
tolerate the error rate no less than that of classical linear codes in classical
situations when the physical resources are very large, namely, $n \to \infty$. In this
case, the classical Gilbert-Varshamov bound is
\begin{equation}
\frac{k}{n}\ge 1-H(\frac{t}{n}),
\end{equation}
where the function $H(x)$ is the Shannon entropy, $H(x)=-x \log_2{x}
-(1-x)\log_2(1-x)$. When the following condition holds,
\begin{equation}\label{correct_condition}
1-H(\frac{t}{n}) > 0,
\end{equation}
there exists a good $[n,k,t]$ classical linear code to correct at most $t$ bit-flip
errors\cite{Good, ClassicalGood}. When finding such $[n,k,t]$ classical code, we can
then work out a corresponding $[[2n,2k,2t]]$ one-party-QECC, which are able to correct
at most $t$ bit-flip and $t$ phase-flip errors on the second halves of the EPR pairs.
The numerical result of condition is $\frac{t}{n}<0.5$. It means that if there are no
more than one half of the EPR pairs corrupted by the errors, there exists a
one-party-QECC which can successfully correct all errors. This result is consistent
with the result in section \ref{s_con}.

To sum up the discussion of this section, we have constructed a generalized series of
the one-party-QECC from the corresponding classical error correcting codes. The
correspondence between our one-party-QECC and the classical linear codes guarantees our
one-party-QECC be able to protect quantum states from at most 50\% bit error rate of
the quantum channel.

\section{Application: quantum dense coding with one-party-QECC}
\label{s_densecoding}

In this section, we present an application of our novel one-party-QECC in quantum
communication: quantum dense coding with one-party-QECC. Quantum dense coding can
transmit more information with a small number of information
carrier\cite{densecoding,longliu}, and has important applications in quantum
communication. Suppose that the server, namely Bob, first sends the second halves of
entangled pairs to a user, namely Alice, and they use other quantum error correction
methods such as EPP to distribute high-fidelity EPR pairs between them. Then, Alice
stores her qubits into the quantum memory for future use. However, as the conditions in
the user's side are always worse than those in the server's side. Alice, the user, may
soon find that her qubits may have more errors, while the qubits in Bob, the server's
side, are still nearly correct. If they want to make use of these impure EPR pairs in
quantum communication, conventionally they should first distill the pairs into nearly
pure using EPP. However, when the error rate in Alice's side is higher than the
threshold of EPP, namely, the fidelity of the impure state is less than the minimum of
0.5 for EPP\cite{EPP1,EPP2}, no successful purification can be done and the state is
totally a mixed state without any entanglement. Are these pairs actually useless?
Indeed, we will demonstrate that Alice and Bob still can make use of the quantum
correlation in the very low fidelity entanglement resources in quantum dense
coding\cite{densecoding,longliu} using one-party-QECC.

First we give a brief introduction of the process of the quantum dense coding with
initial pure EPR pairs and through the noiseless channels. Quantum dense coding makes
use of previous established entangled pairs to send two bits classical information over
only one qubit. Suppose that before the process, Alice and Bob have already shared a
sequence of pure EPR pairs, each in the state of $|\Phi^+\>_{AB}$; in each pair, one
qubit is held by Alice while the other is at the Bob's. Alice wants to transfer some
information to Bob. She makes use of the entangled qubits to send as much as 2 bits of
information with 1 qubit. To be more specific, for each qubit on her side, she can
perform one of the four possible operations: $I(Identity), X, Y, Z$. With simple
calculations, these operations effectively transform the initial state $|\Phi^+\>_{AB}$
into one of the four possible Bell states: $|\Phi^+\>_{AB}, |\Psi^+\>_{AB},
|\Psi^-\>_{AB}, |\Phi^-\>_{AB}$. After encoding all information into the sequence of
the qubits, she sends them to Bob. If there is no error in the channel, after receiving
the qubits from Alice, Bob obtains both halves of the EPR pairs in different Bell
states. Then he performs the Bell measurements which distinguish the four orthogonal
Bell states and retrieves the encoded information from Alice. In summary, the quantum
dense coding employs 1 EPR pair to encode 2 bit information while only 1 qubit travels
through the channel.

Now, let us examine the quantum dense coding in the noisy situation. The protocol
contains two sources of errors. One comes from the initial impure EPR pairs. The
impurity may be brought by the errors in Alice's qubits which can be characterize by
bit-flip error, phase-flip error or both of them. The other comes from the channel
noise. When the quantum channels are noisy, any transmitted qubits may be also altered
by bit-flip error, phase-flip error or both of them. One may consider using EPP to
correct both errors. But if the errors are high enough, EPP will fail. Even if the
channel is noiseless but the fidelity of the initial impure EPR pairs is lower than
0.5, no EPR pairs can be purified and this method fails. However, if we make use of
one-party-QECC, quantum dense coding is still feasible.

To be more specific, to protect the qubit against error from both sources, we should
introduce error correction procedure. Remembering the criteria of the suitable quantum
communications of the one-party-QECC, we learn that if one of the communication parties
finally obtains both halves of the EPR pairs, the home qubits, the joint measurements
and operations in one-party-QECC can be applied. In addition, the different storage
conditions of Alice and Bob and the noisy channels brings much more errors to the
flying qubits sent from Alice to Bob. Thus the errors in the flying and home qubits are
unbalanced. The process of the quantum dense coding that Alice encodes the message into
her qubits and then sends them to Bob who finally holds the whole pairs is suitable for
the application of the one-party-QECC.

With this idea, we transform the quantum dense coding into error-tolerating one.
Normally, Alice and Bob should first share the initial encoding states(not required to
be pure). Then Alice encodes the information by the logical operations and sends them
to Bob. Bob receives the qubits from Alice and measures the stabilizer to obtain the
error syndromes. With the error syndromes, Bob identifies the error pattern and chooses
appropriate recovering operation. Finally he performs the logical measurements on the
logical states and retrieves the message. In a word, all the operations and
measurements are changed into the logical ones. Reviewing section \ref{s_gen}, suppose
that there are $2k$ bit information to be transmitted and $[[2n, 2k, t]]$
one-party-QECC is applied. The initial encoding state is given by
Eq.(\ref{initial_state}), which is equal to $n$ physical EPR pairs each in the state of
$|\Phi^+\>$. Note that the qubits with odd index are held by Alice and those with even
index are held by Bob. Then it comes to the encoding operations,
Eq.(\ref{joint_operator}). At first sight, there is only one kind of logical
operations, $\bar{X}_{2i}$, containing only the operations on the physical qubits in
Alice's (the qubits with odd index). This fact means that Alice would have only two
choices, $I$ and $\bar{X}_{2i}$ to transforms one logical EPR pair. However, there are
two other possible choices corresponding to each $\bar{X}_{2i} = Z_{2i_1-1} Z_{2_i2-1}
\cdots Z_{2i_l-1}$:
\begin{equation}\label{dc_op}
\begin{split}
\bar{U}_{2i} &= X_{2i_1-1} X_{2i_2-1} \cdots X_{2i_l-1},\\
\bar{V}_{2i} &= Y_{2i_1-1} Y_{2i_2-1} \cdots Y_{2i_l-1},\\
\end{split}
\end{equation}
where the subscripts $i_1, i_2, \cdots, i_l$ are the same to those in the
definition of $\bar{X}_{2i-1}$. These $\bar{U}_{2i}$ and $\bar{V}_{2i}$, though
do not commute with the logical operations in Eq.(\ref{joint_operator}),
commute with the stabilizers in Eq.(\ref{joint_stab}) and successfully
transform the initial encoded state. Particularly, $\bar{X}_{2i}$,
$\bar{U}_{2i}$ and $\bar{V}_{2i}$ can transform the encoded state
$|\overline{00}\>_{2i-1, 2i} = |\bar{\Phi}^+\>_{2i-1, 2i}$ to the states
$|\overline{01}\>_{2i-1, 2i} = |\bar{\Phi}^-\>_{2i-1, 2i}$,
$|\overline{10}\>_{2i-1, 2i} = |\bar{\Psi}^+\>_{2i-1, 2i}$ and
$|\overline{11}\>_{2i-1, 2i} = |\bar{\Psi}^-\>_{2i-1, 2i}$ respectively, up to
a global phase. Consequently, Alice still has four choices, $I, \bar{X}_{2i},
\bar{U}_{2i}, \bar{V}_{2i}$ to encode 2 bits information into 1 logical EPR
pair. After encoding the message, Alice sends all her qubits to Bob. Bob
receives them and measures the stabilizers in Eq.(\ref{joint_stab}). If the
bit-flip and phase-flip errors are no more than $t$ respectively on the
transmitted qubits, the measurement results reveal the exact positions of each
error and Bob chooses proper bit-flip and phase-flip operations on those
positions to recover the correct state. With the correct state, Bob performs
the logical Bell measurements, i.e., measures $\bar{X}_{2i-1} \bar{X}_{2i}$ and
$\bar{Z}_{2i-1} \bar{Z}_{2i}$ on the $i$-th logical EPR pair, and fully
retrieves the message of Alice.

Furthermore, as the depolarizing channel in six-state quantum key distribution provides
higher tolerable channel bit error rate\cite{sixstate}, in order to make use of the
same advantage in quantum dense coding with one-party-QECC, we should make sure that
the final bit, phase and both bit and phase errors are symmetric, as an effective
depolarizing channel. Firstly, to deal with the initial impure EPR pairs, we employ the
similar method in Ref.\cite{EPP1} to obtain the pairs with symmetric errors. In
particular, Alice first applies $Y$ to her qubits in order to transform $|\Phi^+\>$
into $|\Psi^-\>$. Then Alice and Bob employ bilateral $\pi/2$ rotations $B_x, B_y,
B_z$\cite{EPP1} to their pairs. These operations are capable of obtaining a
rotationally symmetric mixture, i.e., a Werner state\cite{Werner},
\begin{equation}
W_F = F|\Psi^-\>\<\Psi^-| + \frac{1-F}{3}|\Phi^+\>\<\Phi^+| +
\frac{1-F}{3}|\Phi^-\>\<\Phi^-| + \frac{1-F}{3}|\Psi^+\>\<\Psi^+|,
\end{equation}
where $F$ is the initial fidelity of the impure pairs. Finally, Alice applies
$Y$ to her qubits to revert $|\Psi^-\>$ into $|\Phi^+\>$ and thus obtains a
sequence of impure pairs with symmetric errors.

Secondly, the errors introduced by the noise channels can also be symmetrized.
It can be done by the same method of six-state quantum key
distribution\cite{sixstate}. To be more specific, Alice randomly applies
$I(Identity), T, T^2$ to the qubits after encoding the message, where
\begin{equation}
T = \frac{1}{\sqrt{2}}\left(\begin{array}{cc}\ 1 & \ -i \\ \ 1 & \
i\end{array}\right).
\end{equation}
These operations effectively transform the qubit into $Z$, $X$ and $Y$ basis. When Bob
receives the qubits, Alice tells her operations and then Bob uses $I, T^2, T$ to
recover all the qubits to the $Z$ basis, corresponding to Alice's operations. So with
this manipulation, the three kinds of error probabilities of the quantum channel is
averaged. Thus the final errors that comes from both sources become symmetric. In
summary, the process of the quantum dense coding with one-party-QECC is presented as
following:

\emph{Protocol1: Quantum dense coding with one-party-QECC}
\begin{enumerate}

\item Alice and Bob initially share a group of $n$ impure EPR pairs, with fidelity
$F_0$ to the pure state of $(|\Phi^+\>\<\Phi^+|)^{\otimes n}$.

\item Alice and Bob first make the impure state into a rotational symmetric
mixture. Alice applies $Y$ to her qubits. Then Alice and Bob employ bilateral
$\pi/2$ rotations $B_x, B_y, B_z$ to their pairs. After that, Alice applies $Y$
to her qubits again and the errors of the impure state are symmetrized.

\item Alice wants to transmit $2k$ bit message. She and Bob agree on a $[[2n, 2k, t]]$
one-party-QECC which fully protect the qubits from the errors of the channel.

\item Alice and Bob make use of the shared $n$ EPR pairs from the initial
prepared collection. The initial state is described as the logical state of
$|\underbrace{\overline{00\cdots0}}_{2k}\>$, where the odd qubits are held by
Alice and the even ones are held by Bob.

\item Then Alice encodes the message into the EPR pairs. For the $i$-th
logical pair $|\overline{00}\>_{2i-1, 2i}$, according to the corresponding 2
bit values of the message, Alice chooses one of the four operations, $I,
\bar{X}_{2i}, \bar{U}_{2i}, \bar{V}_{2i}$ to transform the pair into one of the
four states, $|\overline{00}\>_{2i-1, 2i}, |\overline{01}\>_{2i-1, 2i},
|\overline{10}\>_{2i-1, 2i}, |\overline{11}\>_{2i-1, 2i}$.

\item Alice randomly apply $I, T, T^2$ operations to all $n$ qubits on her side, then
sends them to Bob through the noisy quantum channels.

\item Bob receives the qubits and then there are total $2n$ qubits
on his side.

\item Alice tells Bob her choices of $I, T, T^2$ operations.
Accordingly Bob applies $I, T^2, T$ to the received qubits, in order to reverse
Alice's operations. If they are conducting key distribution in this protocol,
they can also random permute the sequence of the EPR pairs and publicly notify
each other.

\item Bob measures the stabilizers of Eq.(\ref{joint_stab}) of the $[[2n, 2k, t]]$
one-party-QECC on the corrupt state. With the measurement results as the error
syndrome, he knows the positions of the errors and applies proper recovering operations
to obtain the correct state.

\item Bob performs the logical Bell measurements on each logical EPR pair and retrieves
the $2k$ bit message.
\end{enumerate}

Finally, we demonstrate the advantages of one-party-QECC in quantum dense coding.
Generally, a noisy quantum channel is defined with three parameters $(p_X, p_Y, p_Z)$
which fully characterize the error probabilities of each qubit. $p_X$ means the
probability of only a single bit-flip error occurring at a qubit which is transmitted
through the noisy quantum channel. $p_Z$ means the probability of only a single
phase-flip error and $p_Y$  the probability of both a bit-flip error and a phase-flip
error. So the fidelity of the channel is $F=1-p_X-p_Y-p_Z$. In the quantum dense coding
above, the effective channel is a depolarizing channel, the three probabilities are
symmetric, namely, $p_X = p_Y = p_Z = p$, and the fidelity is $F = 1-3p$. In section
\ref{s_gen}, we showed that a $[[2n, 2k, t]]$ one-party-QECC can tolerate at most $t$
bit-flip errors and $t$ phase-flip errors. Thus the success of the $[[2n,2k,t]]$
one-party-QECC requires that
\begin{equation}
\begin{split}
\frac{t}{n} \geq p_Z+p_Y = 2p, \\
\frac{t}{n} \geq p_X+p_Y = 2p.
\end{split}
\end{equation}
It follows that
\begin{equation}
F \geq 1 - \frac{3}{2} \frac{t}{n}.
\end{equation}
In asymptotic situation, because $t/n < 0.5$, so $F > 0.25$ and the correctable
fidelity of one-party-QECC reaches 0.25, which is much lower than the lower fidelity
bound of 0.5 in EPP\cite{EPP1, EPP2}. As indicated in previous works, the situation
that the fidelity reaches 0.5 means that the entangled pairs are a total mixture state
without any entanglement, and that no advantage of entanglement is helpful in quantum
communication. However, using one-party-QECC, the quantum dense coding that makes use
of established quantum correlation is still able transmit classical information,
although the fidelity is much lower. Moreover, the correspondence between
one-party-QECC and classical linear codes in section \ref{s_gen} shows that under the
same channel bit error rate, quantum dense coding with one-party-QECC can transmit
classical bit information twice as much as that by classical communication, although
only the same amount of information carriers(qubits and bits) are transmitted. This is
also the result of quantum correlation established prior to the communication. The
application of one-party-QECC also extend the requirement of quantum memory: although
the errors introduced by Alice's quantum memory make the fidelity lower than 0.5 and
forbid the conventional EPP, quantum dense coding with one-party-QECC is still feasible
if the combined errors of both the memory and the channels guarantee the final fidelity
upper to 0.25.

\section{Application: Quantum Secure Direct Communications}\label{s_qsdc}

Based on quantum dense coding, one-party-QECC can be applied to many more protocols due
to the extensive use of dense coding in quantum information processing. In this
section, we present the application of one-party-QECC in a kind of quantum secure
direct communications(QSDC), that are based on quantum dense coding. We also show that
the application can increase the performance of QSDC through noisy quantum channels.

A typical QSDC protocol through noiseless channels consists of two transmission
phases of qubits. In the first phase, Bob creates a sequence of EPR pairs all
in $|\Phi^+\>$ state and distributes the second halves of the pairs to Alice's
side. In the second phase, Alice encodes her message using certain operations
on her received qubits, then sends them back to Bob. Finally, Bob employs Bell
measurements to retrieve Alice's encoded message. As the requirement of
security, Alice and Bob performs error check after both phases. The concrete
protocol is described as:

\emph{Protocol 2: QSDC protocol through noiseless channel}
\begin{enumerate}

\item Bob first prepares a sequence of $3n$ EPR pairs in the state
of $|\Phi^+\>$.

\item Bob chooses a random $3n$ bit binary string $b$, applied
Hadamard transformation $H$ to the second halves of the pairs in which the
corresponding bits of $b$ are 1. Then he sends the second halves to Alice.

\item Alice receives the qubits and publicly acknowledges her
receipt.

\item Alice and Bob randomly chooses an $n$ subset of the EPR
pairs as first-round check pairs. Bob tells Alice the bit values of $b$ in the
place of the check pairs. Then Alice applies $H$ to the qubits of the
first-round check pairs in her part where the corresponding bits of $b$ are 1.
They both measure the check qubits in their halves of the check pairs
respectively in the $Z$-basis. Note that because $H \otimes H = I(Identity)$,
the results of Alice and Bob in each pairs will be the same if there is no
error. Therefore, if they find that there are too many inconsistencies, they
know that the transmitting qubits are eavesdropped and abort the protocol.

\item Alice randomly selects $n$ subset of the rest $2n$ EPR
pairs as second-round check pairs; the rest are served as code pairs.

\item Alice wants to send a $n$ bit binary sequence of message $M$. She encodes
$M$ to the $n$ qubits in her halves of the code pairs by applying $S=-iY=XZ$ to
them where the corresponding bit of $M$ is 1. Then she returns the $2n$ qubits
to Bob.

\item Bob receives the qubits from Alice, applies $H$ to the received qubits where the
corresponding bits of $b$ are 1. Then he publicly announces his receipt.

\item Alice publicly announces the places of second-round check pairs. The same fact
holds that if Bob measures both the qubits in each check pairs in $Z$-basis
respectively, he will get the same results if there is no error. Thus if Bob gets too
many errors, the protocol is aborted.

\item Bob measures both the qubits of the rest $n$ code pairs in $Z$-basis. Because
$S|\Phi^+\> = \frac{1}{\sqrt{2}}(|01\>-|10\>)$ and $(H \otimes S \otimes H)|\Phi^+\> =
-\frac{1}{\sqrt{2}}(|01\>-|10\>)$, the different results of the measurements on the two
qubits in one pair tell Bob that the corresponding bit of $M$ is 1 while the same
results tell that the corresponding bit of $M$ is 0. Therefore, Bob can retrieve the
full information of $M$.
\end{enumerate}

Analyzing protocol 2, we find that the two phases are related to entanglement
distribution and dense coding respectively. In particular, in the first phase,
the sequence Bob creates is a pure state and he sends the second halves of the
EPR pairs to Alice. Thus they are able to perform EPP on the distributed states
in order to correct potential errors. We note that in order to perform EPP, Bob
needs to reveal his random Hadamard transformations on the rest $2n$ pairs to
Alice and thus Alice should perform again the random Hadamard transformations
before the second phase. As $H_1 H_2 |\Phi^+\>_{12} = |\Phi^+\>_{12}$, Bob can
perform the same Hadamard transformations on the qubits in his side before the
EPP and then perform again after the EPP. We also note that the random
selection of first-round check pairs gives the correlation between the check
pairs and the rest pairs, which guarantees that the estimation of error rates
in the rest pairs are of great probability bounded by the error rates in the
check pairs. As a result, they should make sure the estimation of error rates
in the rest pairs does not exceed the capability of EPP. A general EPP with
two-way classical communications is able to correct corrupted EPR pairs with
fidelity $F<0.5$\cite{EPP1, EPP2}. As Bob's random Hadamard transformation
makes the bit error rates and phase error rates symmetric, namely, $p_X+p_Y =
p_Z+p_Y$. Because the check procedure does not distinguish $p_Y$ from $p_X$ or
$p_Z$, the worst case for the errors is $p_Y=0$ and $p_X=p_Z$\cite{twoway}. In
this situation, the channel bit error rate $p_{bit} = p_Z+p_Y = F/2 < 25\%$.
Therefore, Alice and Bob can use EPP to correct errors as long as the tolerable
channel bit error rates $p_{bit}$ is less than $25\%$.

In the second phase, Alice first encodes her message $M$ using $S$ and then sends her
halves of the EPR pairs distributed in the first phase back to Bob. This process is
effectively a dense coding protocol. As shown in section \ref{s_densecoding}, there are
two choices for Bob to correct the errors on the qubits sent by Alice. The first choice
is to use EPP, which is also able to achieve this as long as the channel error rates is
less than $25\%$. On the other hand, if Bob makes use of the alternative solution of
one-party-QECC, he can enhance the performance. The qubits sent from Alice to Bob is
flying qubits, while the qubits on Bob's hand is home qubits. It is also obvious that
Bob in the final stage receives both halves of the EPR pairs. As a result, the
condition for applying one-party-QECC is satisfied. Reviewing section
\ref{s_densecoding}, the operation $S$ Alice uses in encoding is related to the
operation $-iY$, thus the logical operation of Alice is $\bar{S}_{2i} = -i\bar{V}_{2i}$
in Eq. \ref{dc_op}, in correspondence of the $i$-th bit of $M$. In this way,
one-party-QECC can be applied and the detailed protocol through noisy channels is:

\emph{Protocol 3: QSDC protocol with one-party-QECC}
\begin{enumerate}

\item Bob first prepares a sequence of $3n$ EPR pairs in the state
of $|\Phi^+\>$.

\item Bob chooses a random $3n$ bit binary string $b$, applies Hadamard
transformation $H$ to the second halves of the pairs in which the corresponding
bits of $b$ are 1. Then he sends the second halves to Alice.

\item Alice receives the qubits and publicly acknowledges her receipt. Bob
tells Alice the bit values of $b$. Then Alice applies $H$ to the qubits in her
part where the corresponding bits of $b$ are 1.

\item Alice and Bob randomly chooses an $n$ subset of the EPR pairs as first-round
check pairs. They both measure the check qubits in their halves of the check pairs
respectively in the $Z$-basis. Note that because $H \otimes H = I(Identity)$, the
results of Alice and Bob in each pair will be the same if there is no error. Therefore,
if they find that there are too many inconsistencies, they know that the transmitting
qubits are eavesdropped and abort the protocol.

\item Alice and Bob uses a suitable EPP to purify their rest EPR pairs.

\item Alice randomly selects $m$ subset of the rest $2m$ first-level logical EPR
pairs as second-round check pairs; the rest are served as code pairs. She also
randomly chooses a $2m$ bit binary string $b'$, applies first-level logical
Hadamard transformation $\bar{H}$ to the second halves of the pairs in which
the corresponding bits of $b$ are 1. Then he sends the second halves to Alice.

\item Alice wants to send a $k$ bit binary sequence of message $M$. She picks a $[[2m,
2k, t]]$ one-party-QECC that can correct the errors in the second transmission. In the
view of one-party-QECC, there are $k$ second-level logical EPR pairs in the code pairs.
She encodes $M$ to her halves of the second-level logical qubits in the code pairs by
applying $\bar{S}_{2i}=-i\bar{Y}_{2i}$ to them where the corresponding bit of $M$ is 1.
Then she returns all her qubits to Bob.

\item Bob receives the qubits from Alice and publicly announces his receipt.
Then Alice announces $b'$, and Bob applies the first-level logical $\bar{H}$ to
the received first-level qubits where the corresponding bits of $b'$ are 1.

\item Alice publicly announces the places of second-round check pairs and the
one-party-QECC she chooses. The same fact holds that if Bob measures the both qubits in
each check pairs in $Z$-basis respectively, he will get the same results if there is no
error. Thus if Bob gets too many errors, the protocol is aborted.

\item Bob uses the $[[2m,2k,t]]$ one-party-QECC to correct the errors on the rest $m$
first-level logical EPR pairs and obtains $k$ second-level logical code pairs.

\item Bob measures both the qubits of the rest $k$ second-level logical code pairs in
$Z$-basis. Therefore, from the comparison of the measurements on corresponding pairs,
Bob can retrieve the full information of $M$.
\end{enumerate}

The maximal tolerable channel bit error rate in the second phase using one-party-QECC
is obviously $50\%$, according to section \ref{s_con} and \ref{s_gen}. Compared to the
result of $25\%$ for EPP, this result clearly show the enhancement of one-party-QECC in
the application of QSDC.

\section{Conclusion}
\label{s_conclusion}

In this paper, we first present and discuss a novel type of one-party quantum error
correcting codes. We consider a kind of quantum communication with unbalanced errors on
flying and home qubits. In this kind of communication, instead of distributing the
entangled pairs, Bob finally receive both qubits of the entangled pairs. Thus the
feasibility of joint measurements and operations between the flying and home qubits in
error correction enlightens us to introduce them to quantum error correcting code,
called one-party-QECC. Therefore, in section \ref{s_621}, similarly to the classical
[3,1] linear code, we pair up 3 EPR pairs to form a [[6,2,1]] one-party-QECC. Then we
concatenate the [[6,2,1]] one-party-QECC in section \ref{s_con} in order to encode
longer codes. In section \ref{s_gen}, we derive an generalized $[[2n,2k,t]]$
one-party-QECC which covers a large series of the one-party-QECC. Consequently we
establish the formalism of the novel one-party-QECC. At the same time, we analyze the
correcting process of one-party-QECC and derive its tolerable channl bit error rate
50\% which is higher than previous codes. In section \ref{s_densecoding} and
\ref{s_qsdc}, we give an application of one-party-QECC in the quantum dense coding and
quantum secure direct communications and show that they can be conducted with much
higher capacity due to the advantages of one-party-QECC.

Recently, some other proposals on entanglement-assisted quantum error correcting codes
have been discussed\cite{Bowen, catalytic}. They follow the same idea to make use of
entanglement to construct quantum codes. But such codes are partial
entanglement-assisted while our one-party-QECC provides quantum coding from only
entanglement. In addition, we present a physical picture of the one-party-QECC and
thorough analysis of its capacity and applications.

Throughout these sections, two main ideas play an important role. The first one is the
joint measurements and operations between two kinds of qubits, based on the quantum
entanglements and correlations. The second one is the correspondence between the
one-party-QECC and the classical linear codes. In the one-party-QECC, we effectively
treat the EPR pairs as the basic elements with four different values, $|00\>$, $|01\>$,
$|10\>$ and $|11\>$, which play the same role of the bits in the classical codes. The
errors on only one part of each pair causes its value changes rather than the phase
changes, similar to the classical situation. With such correspondence to the classical
linear codes, we easily find the elements of the one-party-QECC, i.e., the stabilizers,
logical states and operations. Moreover, the correspondence guarantees the same high
error-tolerating capability and efficiency of the one-party-QECC to that of the
classical codes, which is the most intriguing result of the discussion.

Moreover, the idea of the one-party-QECC is also valuable to the implementation of
quantum computation. In the one-party-QECC and its applications in quantum
communication, one part of the state, i.e., the home qubits in Bob's hands all through
the protocol, are stored in some storage which provide low error environment, while the
other part of the state, i.e., the flying qubits transmitted from Alice to Bob, are
affected by much more errors in an open environment. It is stated in section
\ref{s_densecoding}, the introduction of one-party-QECC loose the requirement of
quantum memory in Alice, the user's side. Thus, in more general quantum computing
schemes, sometimes there are also unbalanced errors on two different parts of qubits.
We divide these two parts into two layers. In the first layer, the qubits stored in the
storages with very low errors are protected by some error correction scheme, simpler
but tolerating fewer errors. In the second layer, it is feasible to protect the qubits
in much more noisy environments by the one-party-QECC, for in the view of this layer,
the errors only happen in this part of the qubits with much higher error rate than the
part handled in the first hierarchy. Such idea of one-party quantum error correction
may have many other promising applications in the quantum computers with devices of
different error levels.

In conclusion, we establish a series of one-party-QECC and demonstrate its applications
in quantum dense coding and quantum secure direct communications. The one-party-QECC
reveals the amazing strength of the quantum entanglements and joint operations in error
correction. It also brings forth some interesting ideas in designing error correction
schemes. More deep research in the applications of the one-party-QECC in the quantum
communication and computation should be carried out in the future.

\section*{Acknowledgment}
  This work is supported by the 973 Program Grant No. 2006CB921106,
  China National Natural Science Foundation Grant
No. 10325521, 60433050, 60635040,  the SRFDP program of Education Ministry of China,
No. 20060003048 and the Key grant Project of Chinese Ministry of Education No.306020.

\end{document}